\documentclass[twocolumn, prl]{revtex4-1}
\usepackage{graphicx}
\usepackage{epsfig}
\begin{document}
\title{Phase Separation in Mixtures of Repulsive Fermi Gases Driven by Mass Difference }
\author{Xiaoling Cui$^{\dagger, \ast}$ and Tin-Lun Ho$^{\dagger, \ast}$}
\affiliation{$^{\dagger}$ Department of Physics, The Ohio State University, Columbus, OH 43210, USA\\
$^{\ast}$ Institute for Advanced Study, Tsinghua University, Beijing 100084, China}
\date{\today}

\begin{abstract}
We show that phase separation must occur in a mixture of fermions with repulsive interaction if their mass difference is sufficiently large. 
This phenomenon is highly dimension-dependent. Consequently, 
the density profiles of phase separated $3d$ mixtures are very different from those in $1d$. 
Noting that  the ferromagnetic transition of a spin-1/2 repulsive Fermi gas is the equal mass limit of the phase separation in mixtures, we show from  the Bethe Ansatz solution that a ferromagnetic transition will take place in the scattering states when the interaction passes through the strongly repulsive regime and becomes attractive. 
 \end{abstract}

\maketitle

In the last few years, there have been considerable interests in strongly repulsive Fermi gases. Many of these studies were  stimulated by the  initial report of ferromagnetism in the Fermi gas of $^{6}Li$\cite{Ketterle1}.
The possibility of itinerant ferromagnetism was first proposed by Stoner for electron gas\cite{Stoner}. The idea is that if Coulomb repulsion increases faster than kinetic energy with increasing density, as indicated by Hartree-Fock calculation, the system will turn ferromagnetic at sufficiently high densities to avoid  repulsion at the expense of increasing kinetic energy.
However,  Hartree-Fock approximation overestimates repulsion energy. So far, itinerant ferromagnetism has not been found in metals. 

Itinerant ferromagnetism had also been predicted for strongly repulsive Fermi gas based on perturbative and mean field calculations\cite{Allan, Para} prior to the MIT experiment \cite{Ketterle1}. However, such approaches are known to be unreliable in strongly interacting regime. In fact, later experiment has not observed ferromagnetism in strongly interacting $^{6}$Li Fermi gas\cite{Ketterle2}. It is hard to determine whether it is due to the absence of Stoner ferromagnetism or that ferromagnetism is superseded by severe atom loss. 
Still, Stoner's idea of  avoiding repulsion by tuning ferromagnetic remains sound, and should apply  to systems such as Fermi-Fermi mixtures, where the analog of ferromagnetic transition (which leads to magnetic domains) corresponds to phase separation. 

Phase separation of Fermi-Fermi mixtures has been studied in ref.\cite{KC} using  mean field approximation and perturbation methods. It is found that a $^{6}$Li-$^{40}$K mixture will phase separate in the strongly interacting regime. Since mean field theory is know to be unreliable in the strongly interacting regime, 
it raises the questions about whether increasing repulsion can in fact cause a Fermi-Fermi mixture to phase separate. 


In this paper, we would like to point out that phase separation in a Fermi-Fermi mixture can always be induced by increasing the mass ratio of the two fermion species, but not necessarily by increasing repulsion. The reason is that the kinetic energy cost for phase separation can always be reduced to zero by increasing the mass ratio, thereby falling below the repulsion energy, rendering the Stoner argument valid\cite{Jim}.  On the other hand, since the density regime for strong interaction is  dimension dependent, the phenomena of phase separation changes significantly with dimensionality. Since the 
ferromagnetic transition in spin-1/2 systems is the equal mass limit of  the phase separation of Fermi mixtures, it is useful to unify these two phenomena in a global phase diagram as a function of mass ratio and interaction. In the $1d$ case, we  shall also show from exact result that an ``upper-branch" spin-1/2 Fermi gas will turn ferromagnetic as the system passes through the Tonks-Girardeau limit, i.e. 
when the coupling constant jumps from strong repulsion to strong attraction. In the cases we consider, atom loss will not impede the observation of phase separation. 
 

{\em (A). A theorem on mass-difference driven phase separation:} A homogeneous Fermi-Fermi mixture with an arbitrary 
repulsion will  phase separate for sufficiently large mass difference.

First, let us introduce some definitions.  The energy density ${\cal E}_{hm}$ of the ground state of a homogenous mixture of light and heavy fermions with masses ($m_{L}$,  $m_{H}$) and densities ($n_{L}$,  $n_{H}$) is 
\begin{equation} 
{\cal E}_{hm}= {\cal E}_L + {\cal E}_H  + {\cal E}_L 
 G\left( \frac{ m_{L}}{m_{H}} , n_{L}^{1/d} a,  \frac{n_{H}}{n_{L}}\right), 
 \label{mix}\end{equation}
 where ${\cal E}_{L(H)}(n_{L(H)})=A_{d}n^{(2+d)/d}_{L(H)}/m_{L(H)}$ is the energy density of the ideal gas of the light (heavy) fermions, $d$ is the dimensionality, and $A_{d}$ is a constant. The last term $U= {\cal E}_{L} G$ is the interaction energy in units of ${\cal E}_{L}$, and 
  $G$  is a dimensionless function of the variables displayed. $``a"$ is the length scale associated with the interaction. In $3d$, $a$ is the s-wave scattering length $a_s$ in the pseudo-potential $ \hat{U} = 2\pi a_{s}/\overline{m} \sum_{i>j}\delta({\bf r}_{i}-{\bf r}_j)  \left( \frac{\partial}{\partial r_{ij}} r_{ij} \right)$, where $r_{ij}=|{\bf r}_{i}- {\bf r}_{j}|$ for two interacting atoms at ${\bf r}_i$ and ${\bf r}_j$, $\overline{m}^{-1}=m_L^{-1}+m_H^{-1}$, and we have set $\hbar=1$.
By applying harmonic confinement along the axial (with frequency $\omega_z$) or the transverse ($\omega_{\perp}$) direction, the system can be reduced to a quasi $2d$ or a quasi $1d$ system. For quasi $2d$ systems, $a$ is related to the binding energy as $\epsilon_b=1/(2\overline{m}a^2)$, where $\epsilon_b=\frac{A}{\pi} \omega_z e^{\sqrt{2\pi}a_z/a_s}$,  $a_z=\sqrt{1/(\overline{m}\omega_z)}$ is the confinement length and $A\approx 0.915$\cite{Petrov}.  
For quasi $1d$ systems, $a=-\frac{a_{\perp}}{2} (\frac{a_{\perp}}{a_s}-B)$ where $a_{\perp}=\sqrt{1/(\overline{m}\omega_{\perp})}$ and $B\approx1.46$\cite{Olshanii}.  In all dimensions, the energy satisfies the adiabatic theorem, $\partial E_{hm}/\partial \zeta =C/\overline{m}>0$, where $C$ is the contact.  $\zeta$ is  $-1/(2\pi a)$, $\ln (k_0 a) / \pi$ and $a/4$ respectively for 3d, 2d and 1d systems and $k_0$ is an arbitrary momentum scale\cite{Tan,Werner,Zwerger,Molmer}. 
That we parametrize the interaction in terms of $\zeta$ because it is proportional to the magnetic field in experiments that tunes the system across the strongly interacting regime.

\begin{widetext}

\begin{figure}[hbtp]
\includegraphics[height=9cm]{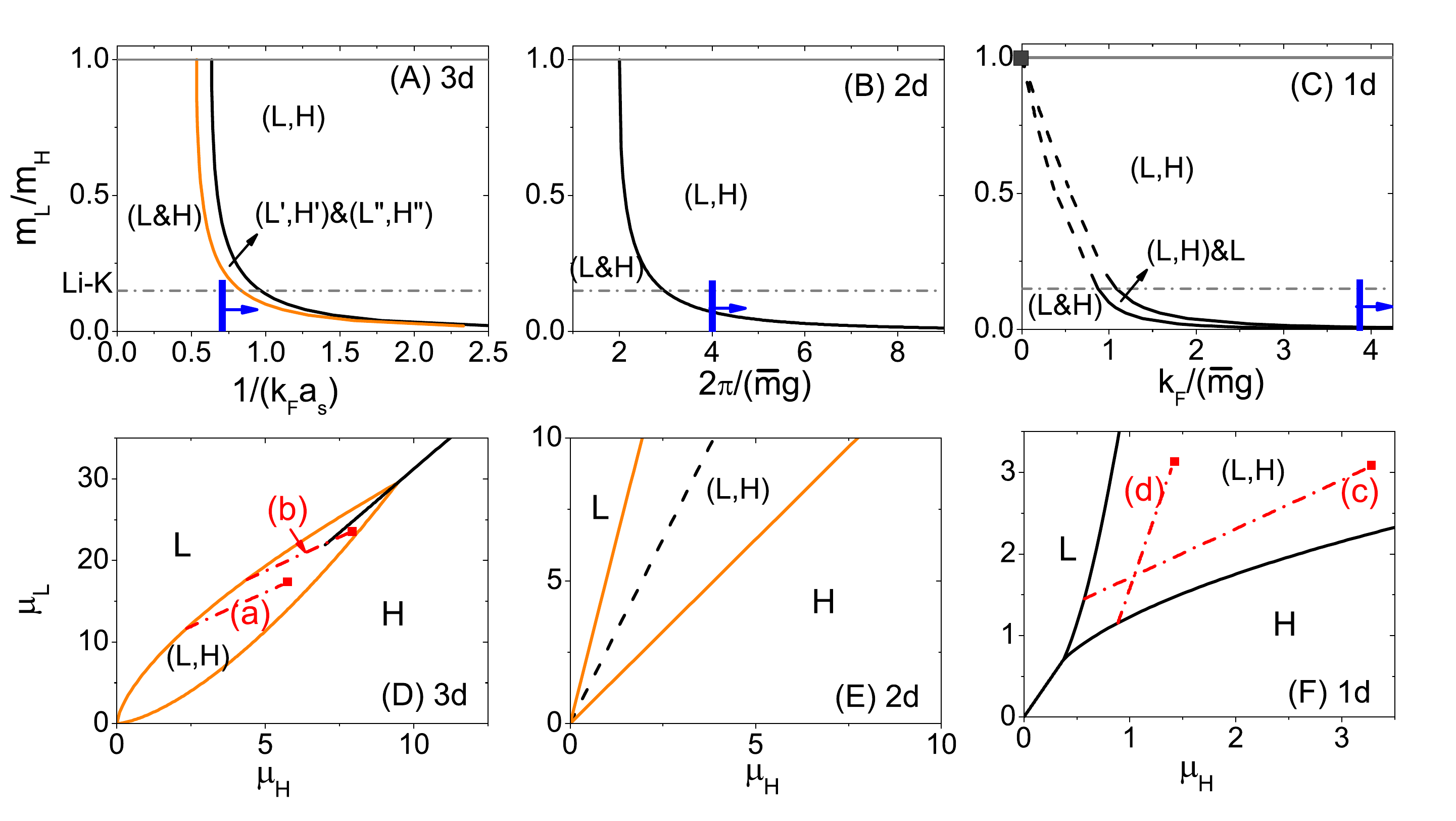}
\caption{Figure 1A, 1B, and 1C are the phase diagrams for a $3d$, $2d$, and $1d$ Fermi-Fermi mixture with $N_{L}=N_{H}$ in a volume $V$. 
$k_F=(6\pi^2 n)^{1/3}$  and $\pi n$ respectively for $3d$ and $1d$, with $n=N_L/V=N_H/V$. To the right of the vertical blue line,  the mean-field interaction energy is less
than half of total kinetic energy for a homogenous mixture (L,H), and the system is weakly interacting deeper in that region. 
The gray dashed-dot lines indicate the case of a Li-K
mixture with $m_L/m_H=6/40$.
Figure 1D, 1E, and 1F are the phase diagrams of a $3d$, $2d$, and $1d$  Li-K mixture in chemical potential plane for weak interactions. 
 $\mu_H,\ \mu_L$ are scaled by
$(\overline{m}^dg^2)^{1/(2-d)}$ in 3d and 1d, and $1/\overline{m}$ in 2d. 
 The red dashed-dot lines in 1D and 1F represent  
trajectories for the density profiles of a trapped system, corresponding to (a-d) in Fig.2, with the squares denoting the chemical potentials at the trap center. 
From Figure 1A to 1F,  the black (orange) solid lines represent the 1st (2nd)-order
boundaries with (without) density discontinuity.  In Figure 1B, the boundary is given by the function $g_c(m_{L}, m_{H}) = 2\pi/\sqrt{m_Lm_H}$. 
In 1E,  the two solid orange lines are the
boundaries for interaction $g<g_c$, with two slopes  $(g/g_c)\sqrt{m_H/m_L}$ and
$(g_c/g)\sqrt{m_H/m_L}$ respectively. When $g\geq g_c$, the two
boundaries merge into one (shown by dashed line) with
slope $\sqrt{m_H/m_L}$. }\label{fig1}
\end{figure}

\end{widetext}

{\em Proof of the Theorem: }
Consider  a system with $N_{L}$ and $N_{H}$ fermions in a  volume $V$, we define 
\begin{equation}
 m_{L}/m_{H}\equiv x, \,\,\,\,\, N_{H}/N_{L} \equiv \gamma, 
 \end{equation}
the total energy of the homogenous mixture  is 
 \begin{equation}
 E_{hm} = V{\cal E}_{L}(n_{L})\left( 1+ \gamma ^{\alpha} x  + G\right), \,\,\,\,\, \alpha=1+2/d. 
 \label{ehm} \end{equation}
Next, we consider the fully phase separated state. Let $V_H$ and $V_{L}$ be the volumes of  the heavy and light fermions, $V_H + V_L = V$.  The ratio $V_H/V_L$ is determined by 
  equating the pressure $P$ of these two separated gases.  Since the pressure of an ideal gas is proportional to its energy density,  $P=2{\cal E}/d$,   we have 
  ${\cal E}_{L}(n'_{L}) = {\cal E}_{H}(n'_{H})$, where  $n'_{H(L)} = N_{H(L)}/V_{H(L)}$. 
  This gives $V_{H}/V_{L} =\gamma x^{1/\alpha}$. 
  The total energy of the phase separated state is 
  $E_{PS} = V_H {\cal E}_{H}(n'_{H}) +   V_L {\cal E}_{L}(n'_{L})
  =   V{\cal E}_{L}(n_{L})(V/V_{L})^{(2+d)/d}$, or
  \begin{equation}
  E_{PS} = V{\cal E}(n_{L})\left( 1+ \gamma x^{1/\alpha} \right)^{\alpha}. 
  \end{equation}
 The phase separated state will have lower energy if $E_{hm} - E_{PS}>0$, or 
 \begin{equation} 
 I(x) = G(x) - \left[  (1+\gamma x^{1/\alpha})^{\alpha}- 1- \gamma^{\alpha} x \right]>0.
\label{condition} \end{equation}
When the mass ratio is sufficiently small such that $\gamma^{1/\alpha} x\ll 1$, hence $x^{\alpha}<x$, Eq.(\ref{condition}) becomes
\begin{equation} 
 I(x) = G(0) - \alpha \gamma x^{1/\alpha} +O(x, x^{2/\alpha}) >0,
 \label{condition2} \end{equation}
where $G(0)>0$ is the repulsive interaction energy in the limit when $m_{H}\rightarrow \infty$\cite{correction}.  Eq.(\ref{condition2}) can always be satisfied for sufficiently small $x$, hence phase separation must occur for sufficiently large  mass difference.  {\em Q.E.D}. 

{\em Corollary:} Because of the adiabatic theorem, if a mixture with mass ratio  $m_{L}/m_{H}$ phase separates at a given interaction parameter $\zeta$, it will continue to phase separate at stronger interactions, i.e. at a larger $\zeta$. 

{\em (B). Phase diagram:}   To demonstrate the effect of mass-imbalance on phase separation, we shall construct the phase diagram as a function of interaction and mass ratio. To obtain results  with certainty, 
we  consider a homogeneous Fermi-Fermi mixture of weakly repulsion. In this case, mean field approach is valid. The energy density ${\cal E}_{hm}$,  the pressure $P$, and the chemical potential $(\mu_{L}, \mu_{H})$ for light and heavy particles are given accurately by 
\begin{eqnarray}
{\cal E}_{hm}(n_{L}, n_{H})&=& {\cal E}_{L}(n_{L})+ {\cal E}_{H}(n_{H}) + gn_{L}n_{H}, \label{E_hm}\\
\mu_{L (H)} (n_{L}, n_{H})&=& \frac{\partial {\cal E}_{L (H)}(n_{L (H)})}{\partial n_{L(H)} }+ gn_{H (L)},
\label{mu-n}\\
P(n_{L}, n_{H})&=&  \mu_{L}n_{L} + \mu_{H}n_{H}- {\cal E}(n_{L}, n_{H}).\label{P} 
\end{eqnarray}
where $g$ is the interaction constant,  $g= \frac{2\pi a_s}{\overline{m}} $ in $3d$, 
$\frac{2\sqrt{\pi}}{\overline{m}}\frac{a_{s}}{a_{z}}$ in quasi $2d$, 
and $\frac{2}{\overline{m}}\frac{a_{s}}{a_{\perp}^2}$ in quasi $1d$. 
While we use the same mean field approach as in ref.\cite{KC}, our ideas are very different.  We goal is to show phase separation must occur at sufficiently large mass ratios, even though the system is weakly interacting. We therefore only draw conclusions in the weakly interacting regime and do not extend our results to strong interacting regions.

To derive the phase diagram,  we consider a system with $N_L$ light fermions and $N_{H}$ heavy fermions in a volume $V$. The possible equilibrium configurations are:  $(a)$ fully phase separated state (PS), denoted as ($L\&H$);  $(b)$ coexistence of a homogenous mixture and a single phase, denoted as $(L,H)\& L$ or 
$(L,H)\& H$; $(c)$  coexistence of two homogeneous mixtures with different densities $(n_{L}', n_{H}')$ and $(n_{L}'', n_{H}'')$, denoted as $(L',H')\&(L'',H'')$;   and $(d)$ a single homogenous mixture $(L,H)$. To determine the presence of these phases, it is sufficient to consider the general case  $(L',H')\&(L'',H'')$,  which covers all  other cases. For example, the state  $(L\&H)$ corresponds to $n_{H}'=n_{L}''=0$. The state $(L,H)\&L$ corresponds to $n_{H}''=0$, and the state $(L,H)$ corresponds to $n_{L}''=n_{H}''=0$. 

Let $(N_{L}', N_{H}')$ and  $(N_{L}'', N_{H}'')$ be  particle numbers  of the mixtures  $(L',H')$ and $(L'',H'')$, and  $V'$ and $V''$ be their volumes respectively. 
The equilibrium configuration is obtained by minimizing the total energy  with 
respect to these particle numbers and volumes, subject to the constraint  $N_{L}'+ N_{L}'' = N_{L}$, $N_{H}'+ N_{H}'' = N_{H}$; and $V'+V''=V$.  The evolution of this equilibrium state as a function of mass ratio and interaction strength yields the phase diagram. 
Figure 1A, 1B, and 1C show the phase diagrams for a $3d$, $2d$, and $1d$ mixture with $N_{L}=N_{H}$ in a volume $V$. 
For both $1d$ and $3d$, there is a range of mass ratio (for given interaction) in which the system consists of two different phases in equilibrium, 
($(L,H)\& L$ for $1d$ and $(L',H')\& (L'', H'')$ for $3d$). This feature is absent in $2d$\cite{polaron}. For all dimenson,  the system is fully phase separated in the weakly interacting regime for sufficiently large mass difference. In this regime, atom loss will be strongly suppressed\cite{Petrov03} and will not hinder the observation of Stoner instability.

Note that the phase boundaries  shown in Figure 1A to 1C are  inaccurate in the strongly interacting region, since they are derived from the mean field expressions Eqns.(\ref{E_hm}), (\ref{mu-n}) and (\ref{P}). 
However,  the corollary in section $(A)$ guarantees that the system will phase separate  in the strongly interacting regime over a range of mass ratio wider than that in the  weakly interacting regime. 

{\em (C) Ferromagnetic transition of  $1d$ spin-1/2 Fermi gas:} The phase diagram for $1d$ Fermi-Fermi mixture is not only constraint by the results in the weakly interacting regime,  but also by the exact Bethe Ansatz solution along the line  $m_{L}/m_{H}=1$\cite{Chen}, which is a  spin-1/2 repulsive Fermi gas with interaction $g\sum_{i>j}\delta(x_i - x_j)$, where  $g=-4(\overline{m} \zeta)^{-1}$. Because of the integrability of this system, there are two classes of eigenstates:  one where all quasi-momenta are real, i.e., all particles are in scattering states, (denoted as class (i)),  and one that contains at least one pair complex conjugate quasi-momenta, i.e. with at least one fermion bound pair, (denoted as class (ii)). Repulsive Fermi gas, which falls into class (i), is referred to as in the ``upper branch";
since it is a many-body {\em eigenstate}, it will not decay into class (ii)\cite{footnote}. 

Experimentally, one can tune the system from weak to strong repulsion ($\zeta=0^{-}, \ g^{-1}=0^{+}$), 
and then to strongly attraction ($\zeta=0^{+}, \ g^{-1}=0^{-}$).  The regime where $g^{-1}=0^{+}$ will be referred to as the Tonk-Girardeau (TG) regime. 
The ground state of a repulsive  ($\zeta<0$ ) spin-1/2 Fermi gas with equal spin population is a spin-singlet according to the Lieb-Mattis theorem\cite{Lieb-Mattis}. 
In the TG limit, the spatial wavefunction of the ground state is
 identical to that of a fully spin polarized Fermi gas up to a sign (which changes in various regions in configuration space).  As a result, its energy $E(0)$ is given by that of a fully spin polarized state with 
  huge spin degeneracy\cite{Chen}  -- all spin configurations including the spin configurations (a) to (c) mentioned above are degenerate, with $H$ and $L$ now labeling the two spin species.
This means that  the two phase boundaries in Fig.1C will converge to the equal mass point $m_{L}/m_{H}= 1$ at resonance.  
Crossing the TG limit to the attractive side, the energies of all spin states continue to increase according to the adiabatic theorem, hence $E(\zeta >0) > E(0)$; except for the largest spin state 
which remains at  $E(0)$ regardless of interaction. As a result, the system will make transition to this maximum spin state.  In practice, such transition can be facilitated by the presence of small magnetic field gradients that destroy spin conservation.  It is useful to note that atom loss in the TG regime is vanishing small\cite{footnoteTG}, and therefore will not affect the observation of ferromagnetism.

\begin{figure}[hbtp]
\includegraphics[height=8cm]{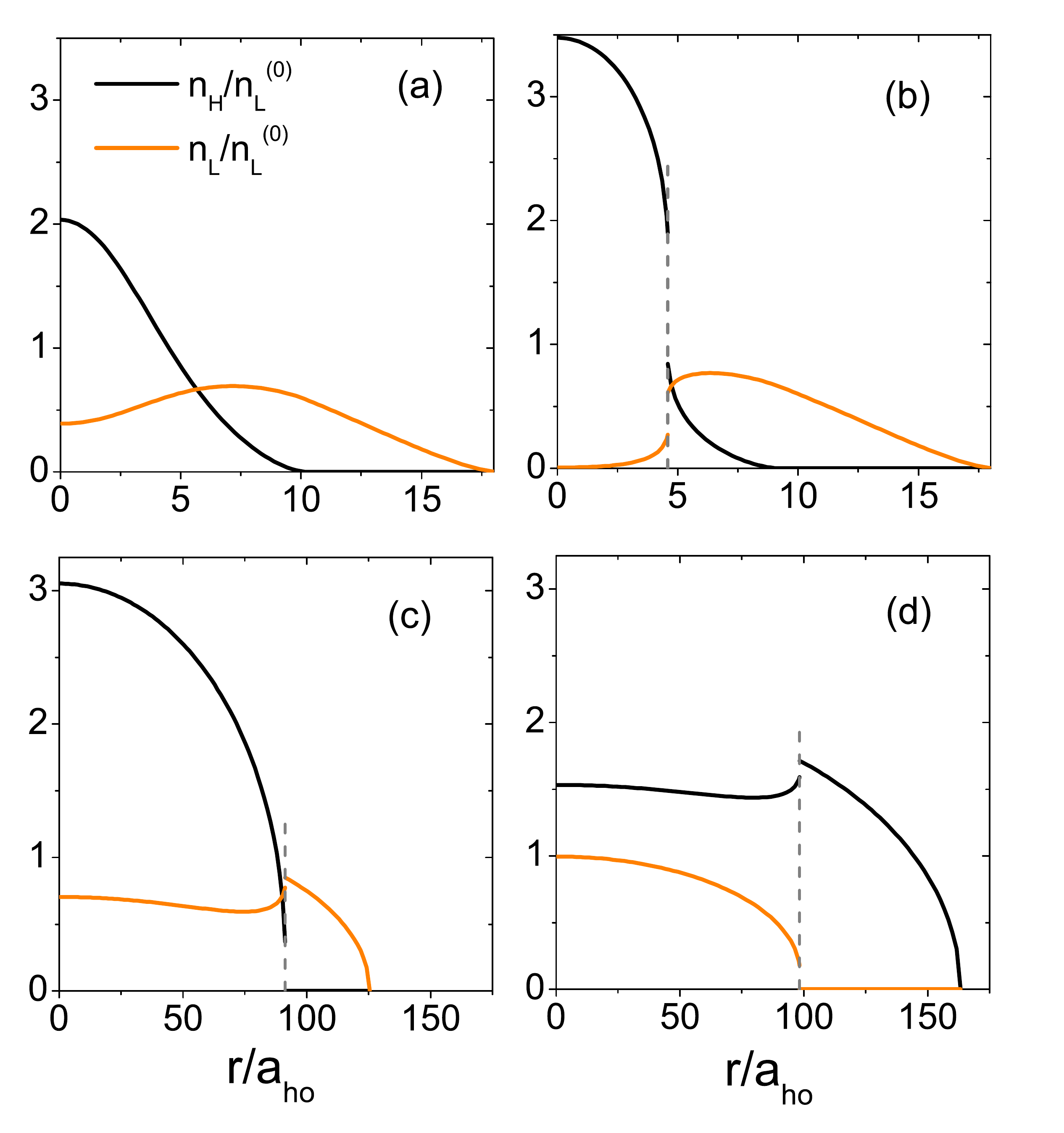}
\caption{Density profiles of a trapped Li(Light)-K(Heavy) mixture in 3d((a) and (b)) and 1d((c) and (d)), corresponding  to the trajectories (a) to (d) in 
Fig.1D and 1F. The densities ($n_H,n_L$) have been normalized by
$n_L^{(0)}$, the density of light atoms(Li) at the trap center
for non-interacting system in the same setup. The position $r$ is scaled by
$a_{ho}=\sqrt{1/(m_L\omega_L)}$, the confinement length of light
atoms.  (a) and (b) are with the same particle numbers
$(N_H,N_L)=(10^5)(1.47,6.9)$ and the same trapping frequency ratio
$\omega_H/\omega_L=0.3$, but with different interaction strengths
$a/a_{ho}=0.055(a),0.065(b)$. (c) and (d) are with the same $(N_H,N_L)=(10^4)(1.45,0.51)$ and the same interaction $-a_{ho}/a=15\pi$, but with different $\omega_H/\omega_L=0.5(c),0.2(d)$. \\
}
\label{fig2}
\end{figure}

 {\em (D) The density profile in a trap:} The density profiles of heavy and light atoms in a trap can be obtained from the equation of state $n_{L (H)}(\mu_{L}, \mu_{H})$ using standard local density approximation (LDA).  Since the equation of state depends on the nature of the equilibrium phase, one needs to first determine its nature as a function of chemical potentials $(\mu_{L}, \mu_{H})$. 

For given ($\mu_{L}, \mu_{H}$), three  phases are possible : the single component Fermi gas $(L)$, $(H)$, and the homogenous mixture $(L,H)$.  
To express the pressure of a homogenous mixture $P_{hm}$ as a function of $\mu_{L}, \mu_{H}$, we invert Eq.(\ref{mu-n}) to obtain $n_{L}$ and $n_{H}$ as a function of $\mu_{L}$ and $\mu_{H}$, and then substitute them into Eq.(\ref{P}).  The pressure of $(L)$ or
$(H)$ is $P_{L(H)}(\mu_L, \mu_{H}) =  B_{d}  m_{L(H)}^{d/2}\mu_{L(H)}^{1+d/2}$,  where $B_d$ is a constant. 
The phase boundary for the full phase separation is  $P_{L}(\mu_L) = P_{H}(\mu_{H})$, or
\begin{equation}
\mu_{H}/\mu_{L}=\beta,\ \ \ \ \ \ \ \ \beta= (m_{L}/m_{H})^{d/(d+2)}.
\label{ps} \end{equation}
The phase boundary between the mixture $(L,H)$ and $L$ (or $(H)$) is obtained by equating $P_{hm}(\mu_{L}, \mu_{H}) = P_{L(H)}(\mu_{L(H)})$. 
The phase boundaries for the $3d$, $2d$, and {\bf $1d$} mixtures are shown in  Figure 1D, 1E, and 1F respectively.  Within the region of homogenous mixture, the inversion of  Eq.(\ref{mu-n}) may yield several solutions of densities (say, $(n_{L}', n_{H}')$, $(n_{L}'', n_{H}'')$) for given chemical potentials $(\mu_{L}, \mu_{H})$. The thermodynamic state is given by the one with highest pressure.  In the $3d$ case, the homogeneous mixture is contained within the ``bubble" in  Figure 1D. Within this region,  the thermodynamic state is unique except on the line that is an extension of the boundary  Eq.(\ref{ps}) where two states (with densities  $(n_{L}', n_{H}')$, $(n_{L}'', n_{H}'')$) have identical chemical potential and pressure.  This is a line of first order transition. 
Furthermore, the densities of these two phases are related as $n_{L}'= \beta n_{H}''$, $n_{H}'= \beta^{-1} n_{L}$, since Eq.(\ref{E_hm}) to (\ref{P})) are invariant under this change.   The density discontinuities  across this line $\Delta n_{L} = \beta n_{H} - n_{L}$,   $\Delta n_{H} =\beta^{-1}n_{L} - n_{H}$ then has the ratio $\Delta n_{L}/\Delta n_{H} = -  \beta$. 

In Fig.2a to 2d, we show the density profiles of the $3d$ and $1d$ mixtures in a trap obtained by 
applying LDA to the equation of state $n_{L(H)}({\bf r}) = n_{L(H)}(\mu_{L}-V_{L}({\bf r}), \mu_{H}-V_{H}({\bf r}) )$, where $V_{L(H)}({\bf r})=m_{L(H)}\omega_{L(H)}^2 {\bf r}^2/2$ are the harmonic potentials experienced by the light(L) and heavy(H) particles. Moving from the center of the trap to the surface of the cloud corresponds to following the trajectories indicated in Fig.1D and 1F.   Fig.2a and 2b show the density profiles of a $3d$ mixture at different interaction strengths. The discontinuities in 
the densities obey the related mentioned above.  Fig.2c and 2d show a $1d$ mixture under different trapping potentials.

Two features of the density profiles should be emphasized. Firstly, the density profiles of a $3d$ mixture differ significantly from that of the $1d$ mixture, (see Fig.1D and 1F). 
Phase separation takes place in the outer part of the atom cloud in $1d$ but in the inner part in $3d$.  This is because the strongly interacting regime occurs in the low (high) density region in $1d$ ($3d$).  Secondly, in Fig. 2a-2d, we note that $n_{L(H)}$ can increase with $r$. This is different from 
the single component case, where $dn/dr<0$, due to the fact that  $dn/d\mu >0$ as demanded by thermodynamic stability.  
 In the mixture case, stability against density fluctuation requires Det$(M) >0$, where $M_{ij} =  \partial \mu_{i}/ \partial n_{j} $, 
and $i, j = L$ and $H$.  We then have $dn_{i}/dr= ( M^{-1})_{ij}d\mu_{j}/dr$, where 
$M^{-1}= {\rm Det}^{-1}(M)\left( \begin{array}{cc}  A_{H} & -g\\-g & A_{L}\end{array}\right)$, 
$A_{L(H)}  = \frac{  \partial \mu_{L(H)}  }{  \partial n_{L(H)}}>0$.  That $d n_{L(H)}/d r$ can be positive or negative is because it is made up of two terms.  If $dn_L/dr>0$, it is easily shown from stability condition ($A_LA_H>g^2$) that $dn_H/dr<0$. Thus one can have at most one species with a positive density derivative.


 {\em Conclusion.} We have shown that the Stoner instability (phase separation) can be driven by large mass difference of Fermi-Fermi mixtures, but not necessarily by strong repulsions. In all dimensions, phase separation will occur for sufficiently large mass difference even in the weak interacting regime. Furthermore, we point out that the Bethe Ansatz solution implies a Stoner instability of the  1d spin-$1/2$ fermions across the TG limit, which inn turn allows one to constrain the phase diagram of 1d Fermi-Fermi mixtures. 
In the cases we consider,   atom loss would be suppressed and will not affect observation of Stoner ferromagnetism in experiments.

XC acknowledges the support of NSFC under Grant No. 11104158, and Tsinghua University Initiative Scientific Research Program. TLH acknowledges the support by NSF Grant DMR-0907366 and by DARPA under the Army Research Office Grant Nos. W911NF-07-1-0464, W911NF0710576, by the Institute for Advanced Study of Tsinghua University through the Qian-Ren Program.


\begin{thebibliography}{99}
\bibitem{correction} Since $\overline{m}$ enters ${\cal E}$ in the combination $a_s/\overline{m}$, we have $\partial {\cal E}/\partial \overline{m} <0$. 
Combining with Eq.(\ref{ehm}), we have $G(x)>G(0)-\gamma^{\alpha} x$, which in turn implies 
the correction to linear order in $x$ in Eq.(\ref{condition2}) vanishes.
\bibitem{footnoteTG} In the TG limit, the atom loss of two-component fermions is strongly suppressed as in hard-core bosons\cite{Gora, Haller11}, as both systems behave equivalent to spinless fermions.
\end{thebibliography}
\end{document}